\def\rfr#1{eq. (\ref{#1})}

\def\derp#1#2{\rp{\partial{#1}}{\partial{#2}}}
\def\dert#1#2{\frac{{{d}}{#1}}{{{d}}{#2}}}              
\def\virg#1{``#1''}

\def\eqi{\begin{equation}}
\def\eqf{\end{equation}}
\def\eqia{\begin{eqnarray}}
\def\eqfa{\end{eqnarray}}
\def\rp#1#2{{#1\over#2}} \def\lb#1{\label{#1}}
\def\bb#1#2#3{\bibitem[\protect\citeauthoryear{#1}{#2}]{#3}}
\def\kx{\hat{k}_x}
\def\ky{\hat{k}_y}
\def\kz{\hat{k}_z}
\def\si{\sin I}
\def\ci{\cos I}
\def\so{\sin\omega}
\def\co{\cos\omega}
\def\sO{\sin\Omega}
\def\cO{\cos\Omega}
\def\su{\sin u}
\def\cu{\cos u}
\def\bds#1{\boldsymbol{#1}}

\documentclass{aastex}
\usepackage{url}
\usepackage{amsmath,amsthm,amscd,amssymb}
\usepackage{latexsym,wasysym}
\usepackage{graphicx,epsfig}

\RequirePackage{color}

\newcommand{\emaila}{lorenzo.iorio@libero.it}

\begin{document}

\title{Orbital effects of spatial variations of fundamental coupling constants}
\shortauthors{L. Iorio}

\author{Lorenzo Iorio\altaffilmark{1} }
\affil{Ministero dell'Istruzione, dell'Universit\`{a} e della Ricerca (M.I.U.R.)-Istruzione. Institute for Theoretical Physics and
High Mathematics Einstein-Galilei. Permanent address for correspondence: Viale Unit\`{a} di Italia 68, 70125, Bari (BA), Italy.}

\email{\emaila}

\begin{abstract}
We deal with the  effects induced on the orbit of a test particle revolving around a central body by putative spatial variations of dimensionless fundamental coupling constants $\zeta$. In particular, we assume a dipole gradient for $\zeta(\bds r)/\overline{\zeta}$ along a generic direction $\bds{\hat{k}}$ in space.
We analytically work out the long-term variations  of all the six standard osculating Keplerian orbital elements parameterizing the orbit of a test particle in a gravitationally bound two-body system. Apart from the semi-major axis $a$, the eccentricity $e$, the inclination $I$, the longitude of the ascending node $\Omega$, the longitude of pericenter $\varpi$ and the mean anomaly $\mathcal{M}$ undergo non-zero long-term changes. By using the usual decomposition along the radial ($R$), transverse ($T$) and normal ($N$) directions, we also analytically work out the long-term changes $\Delta R,\Delta T,\Delta N$ and $\Delta v_R,\Delta v_T,\Delta v_N$ experienced by the position and the velocity vectors $\bds r$ and $\bds v$ of the test particle. Apart from $\Delta N$, all the other five shifts do not vanish over one full orbital revolution. In the calculation we do not use \textit{a-priori} simplifying assumptions concerning $e$ and $I$. Thus, our results are valid for a generic orbital geometry; moreover, they hold for any gradient direction $\bds{\hat{k}}$.
We compare our predictions to the latest observational results for some of the major bodies of the solar system. The largest predicted planetary perihelion precessions occur for the rocky planets, amounting to some $10^{-2}-10^{-3}$ milliarcseconds per century. Apart from the Earth, they are $2-3$ orders of magnitude smaller than the present-day accuracy in empirically determining the corrections $\Delta\dot\varpi$ to the standard Newtonian-Einsteinian planetary perihelion rates. Numerically integrated time series of the interplanetary range for some Earth-planet pairs yield Stark-like signatures as large as $0.1-10$ millimeters; future planned planetary laser ranging facilities should be accurate at a cm level. The long-term variations of the lunar eccentricity and perigee are of the order of $10^{-14}$ yr$^{-1}$ and $10^{-4}$ milliarcseconds per year, respectively, while the change $\Delta R$  in the  radial component of the Moon's geocentric orbit is as large as $0.8$ microns per orbit. A numerically calculated geocentric lunar range time series has a maximum nominal peak-to-peak amplitude of just a few millimeters, with an average of $0.3$ microns over 30 yr. The present-day accuracies in determining $\dot e$ and $\dot\varpi$ for the Moon are  $10^{-12}$ yr$^{-1}$ and $10^{-1}$ milliarcseconds per year, respectively. The APOLLO facility should be able to determine on a continuous basis the Earth-Moon range with a millimeter accuracy.
 \end{abstract}


\section{Introduction}
Testing constancy of the fundamental constants entering the basic laws of physics is of the utmost importance for our understanding of the nature of the gravitational interaction and of the domain of validity of the Einsteinian general theory of relativity; for a recent review see \citet{Uzan} and references therein.

Since the matter-energy content $U=mc^2$ of material bodies generally depends on the parameters  of the Standard Model,  a spatial variation in one of them will induce an extra-force on a body of mass $m$
\eqi\bds F=-\bds\nabla U=-c^2\left(\derp{m}{\zeta}\right)\bds\nabla\zeta,\eqf
where\footnote{The dimensionless ratios of various fundamental parameters are dubbed $r_i, i=1,2,...$ by \citet{DamDon}.} $\zeta$ is an adimensional fundamental parameter like, e.g., the fine structure constant or the electron-to-proton mass ratio, and $c$ is the speed of light in vacuum.
 In particular, for a dipole-type spatial variation \citep{DamDon} \eqi\rp{\zeta(\bds r)}{\overline{\zeta}}=1+B\left(\bds{\hat{k}}\bds\cdot\bds r\right)\lb{dipo}\eqf of $\zeta$, the  force is \citep{DamDon}
 \eqi \bds F=-mQBc^2\bds{\hat{k}},\lb{forza}\eqf
  in which
\eqi Q\doteq\rp{\zeta}{m}\derp{m}{\zeta} \eqf is a dimensionless \virg{charge}.
In \rfr{dipo} and \rfr{forza} $B$ is a slope parameter, having dimensions of L$^{-1}$,  relative to a direction in the space determined by the unit vector $\bds{\hat{k}}$. For example, for the same direction\footnote{It corresponds to equatorial coordinates RA$=17.\textcolor{black}{4}\pm 0.6$ hr, DEC$=-\textcolor{black}{58}\pm \textcolor{black}{6}$ deg \citep{uno}.}
\eqi\textcolor{black}{\bds{\hat{k}}=\{0.50\pm 0.08, -0.19\pm 0.04,-0.84\pm 0.04\}\lb{kappa}}\eqf
with respect to an ecliptic frame, it was found \citep{uno}
\eqi B=(1.10\pm 0.25)\times 10^{-6}\ {\rm Glyr}^{-1}=(1.16\pm 0.26)\times 10^{-31}\ {\rm m}^{-1}\lb{fina}\eqf
for the fine structure constant, and \citep{due}
\eqi B=(2.6\pm 1.3)\times 10^{-6}\ {\rm Glyr}^{-1}=(2.7\pm 1.3)\times 10^{-31}\ {\rm m}^{-1}\eqf
for the electron-to-proton mass ratio.

If $Q$ is not the same for all bodies, then a non-zero, net relative  acceleration of Stark-type
\eqi\bds A\doteq \bds{A}_{\rm B}-\bds{A}_{\rm A}=-\Delta Q B c^2 \bds{\hat{k}},\lb{maronna}\eqf
where
\eqi\Delta Q\doteq Q_{\rm B}-Q_{\rm A},\eqf
occurs for a two-body system A-B. Notice that \rfr{maronna}, which implies a violation of the equivalence principle, holds for a generic adimensional parameter $\zeta$; in principle, the total extra-acceleration is the sum of all the terms like \rfr{maronna} due to the gradients of the various $\zeta$. As far as the magnitude of \rfr{maronna} is concerned, it is certainly quite small. Indeed, from, say, \rfr{fina} it is
\eqi A\lesssim 10^{-14}\ {\rm m\ s}^{-2};\lb{kacc}\eqf suffices it to say that the standard Newtonian inverse-square law for the Sun-Earth system yields
\eqi A^{\oplus}_{\rm Newton}\simeq 10^{-3}\ {\rm m\ s}^{2}.\lb{Newtonacc}\eqf

The plan of the paper is as follows. In Section \ref{calcolo} we analytically work out the long-term effects caused by a Stark-type extra-acceleration of the form of \rfr{maronna} on the motion of a test particle orbiting a central body. We do not make any \textit{a-priori} assumption concerning both the particle's orbital geometry and the direction $\bds{\hat{k}}$. Consequences of a violation of the equivalence principle referred to a fixed direction in space were investigated by \citet{DamScia} in the framework of binary pulsars, and preliminarily by \citet{DamDon} for the Earth-Moon system. In Section \ref{confronto} we compare our results to the latest empirical determinations from solar system observations. Section \ref{conclusioni} is devoted to the summarizing our results.
\section{Analytical calculation of the orbital effects}\lb{calcolo}
The standard Keplerian orbital elements of the orbit of a test particle are the semi-major axis $a$, the eccentricity $e$, the inclination $I$, the longitude of the ascending node $\Omega$, the argument of pericenter $\omega$, and the mean anomaly $\mathcal{M}$. While $a$ and $e$ determine the size and the shape\footnote{The eccentricity $e$ is a numerical parameter for which $0\leq e < 1$ holds; $e=0$ corresponds to a circle.}, respectively, of the Keplerian ellipse, $I,\Omega,\omega$ fix its spatial orientation. $I$ is the inclination of the orbital plane to the reference $\{x,y\}$ plane, while $\Omega$ is an angle in the $\{x,y\}$ plane counted from a reference $x$ direction to the line of the nodes, which is the intersection of the orbital plane with the $\{x,y\}$ plane. The angle $\omega$ lies in the orbital plane: it is counted from the line of the nodes to the pericenter, which is the point of closest approach of the test particle to the primary. In planetary data reduction the longitude of the pericenter $\varpi\doteq \Omega+\omega$ is customarily used \textcolor{black}{by astronomers, although it is a \virg{dogleg} angle}. The argument of latitude $u\doteq \omega+f$ is an angle in the orbital plane which reckons the instantaneous position of the test particle along its orbit with respect to the line of the nodes: $f$ is the time-dependent true anomaly. The mean anomaly is defined as \eqi\mathcal{M}\doteq n(t-t_p),\lb{manol}\eqf where \eqi n\doteq \sqrt{GM/a^3}\lb{meanmo}\eqf is the Keplerian mean motion related to the Keplerian orbital period by $n=2\pi/P_{\rm b}$, and $t_p$ is the time of passage at the pericenter.
In the unperturbed two-body pointlike   case,
 the Keplerian ellipse, characterized by
\begin{equation}
\left\{
\begin{array}{lll}
x &=& r\left(\cos\Omega \cos u - \sin\Omega \sin u \cos I\right), \\ \\
y &=& r\left(\sin\Omega \cos u + \cos\Omega \sin u \cos I\right), \\ \\
z &=& r\left(\sin u \sin I\right),
\end{array}\lb{xyz}
\right.
\end{equation}
and
\eqi r=\rp{a(1-e^2)}{1+e\cos f},\eqf neither varies its shape nor its size; its orientation is fixed in space as well.

A small\footnote{Cfr. \rfr{kacc} and \rfr{Newtonacc}, and the figures for $\Delta Q$ in Table \ref{tavola} below.} perturbing acceleration $\bds A$\textcolor{black}{, like  \rfr{maronna},} of the \textcolor{black}{dominant inverse-square} Newtonian \textcolor{black}{term $A_{\rm Newton}$} induces slow temporal changes of the osculating Keplerian orbital elements.
The Gauss equations for their variation, valid for any kind of acceleration whatever its physical origin may be,  are \citep{BeFa}
\begin{equation}
\left\{
\begin{array}{lll}
\dert a t & = & \rp{2}{n\sqrt{1-e^2}} \left[e A_R\sin f +A_{T}\left(\rp{p}{r}\right)\right],\\   \\
\dert e t  & = & \rp{\sqrt{1-e^2}}{na}\left\{A_R\sin f + A_{T}\left[\cos f + \rp{1}{e}\left(1 - \rp{r}{a}\right)\right]\right\},\\  \\
\dert I t & = & \rp{1}{na\sqrt{1-e^2}}A_N\left(\rp{r}{a}\right)\cos u,\\   \\
\dert\Omega t & = & \rp{1}{na\sin I\sqrt{1-e^2}}A_N\left(\rp{r}{a}\right)\sin u,\\    \\
\dert\varpi t & = &\rp{\sqrt{1-e^2}}{nae}\left[-A_R\cos f + A_{T}\left(1+\rp{r}{p}\right)\sin f\right]+2\sin^2\left(\rp{I}{2}\right)\dert\Omega t,\\   \\
\dert {\mathcal{M}} t & = & n - \rp{2}{na} A_R\left(\rp{r}{a}\right) -\sqrt{1-e^2}\left(\dert\omega t + \cos I \dert\Omega t\right).
\end{array}\lb{Gauss}
\right.
\end{equation}
 In \rfr{Gauss} $p\doteq a(1-e^2)$ is the semi-latus rectum, and $A_R,A_T,A_N$ are the radial, transverse and out-of-plane components of the disturbing acceleration $\bds A$, respectively. In a typical first-order perturbative calculation, they have to be computed onto the unperturbed Keplerian ellipse\footnote{In principle, one may choose a different reference orbit including general relativity itself \citep{Calura1,Calura2}. However, completely negligible mixed Einstein-Stark terms would result. We will not deal with them. Such a conclusion will be \textit{a-posteriori} confirmed by the extremely small magnitudes of the presently computed Newton-Stark effects (See Section \ref{confronto} below).} according to
 \begin{equation}
\left\{
\begin{array}{lll}
A_R & = &\bds A\bds\cdot\bds{\hat{R}}, \\ \\
A_T & = &\bds A\bds\cdot\bds{\hat{T}}, \\ \\
A_N & = & \bds A\bds\cdot\bds{\hat{N}},
\end{array}
\right.
\end{equation}
where the unit vectors along the radial, transverse and out-of-plane directions are \begin{equation}
\bds{\hat{R}}=\left\{
\begin{array}{lll}
\cos\Omega \cos u-\cos I\sin\Omega\sin u, \\ \\
\sin\Omega\cos u+\cos I\cos\Omega\sin u, \\ \\
\sin I\sin u,
\end{array}\lb{erre}
\right.
\end{equation}
\begin{equation}
\bds{\hat{T}}=\left\{
\begin{array}{lll}
-\cos\Omega\sin u -\cos I\sin\Omega\cos u, \\ \\
-\sin\Omega\sin u+\cos I\cos\Omega\cos u, \\ \\
 \sin I\cos u,
\end{array}\lb{tau}
\right.
\end{equation}
\begin{equation}
\bds{\hat{N}}=\left\{
\begin{array}{lll}
\sin I\sin\Omega, \\ \\
-\sin I\cos\Omega, \\ \\
 \cos I.
\end{array}\lb{nu}
\right.
\end{equation}
\textcolor{black}{
The result is, after some algebra,
\begin{equation}
\left\{
\begin{array}{lll}
A_R & = & Bc^2\Delta Q\left\{-\cu\left(\kx\cO+\ky\sO\right)-\su\left[\kz\si+\ci\left(\ky\cO-\kx\sO\right)\right]\right\}, \\ \\
A_T & = & Bc^2\Delta Q\left[-\kz\cu\si -\ci\cu\left(\ky\cO-\kx\sO\right)+\su\left(\kx\cO+\ky\sO\right)\right], \\ \\
A_N & = & -Bc^2\Delta Q\left[\kz\ci +\si\left(\kx\sO-\ky\cO\right)\right].
\end{array}\lb{ARATAN}
\right.
\end{equation}
}

 In the case of \rfr{maronna}, it turns out that it is computationally  more convenient to use the eccentric anomaly $E$ instead of the true anomaly $f$. Basically, $E$ can be regarded as
a parametrization of the usual polar angle $\theta$ in the orbital plane, being defined as \eqi\mathcal{M}\doteq E-e\sin E.\eqf To this aim, useful conversion relations\textcolor{black}{, to be used when \rfr{ARATAN} is inserted in the right-hand-sides of \rfr{Gauss},} are \citep{BeFa}
 \eqi
 \left\{
 \begin{array}{lll}
 \cos f &=& \rp{\cos E-e}{1-e\cos E}, \\ \\
 \sin f &=& \rp{\sqrt{1-e^2}\sin E}{1-e\cos E},\\ \\
 r &=& a(1-e\cos E), \\ \\
 dt &=& \left(\rp{1-e\cos E}{n}\right)d E.
 \end{array}\lb{conz}
 \right.
 \eqf
 \textcolor{black}{Cumbersome calculation yield} the  variations of all the Keplerian osculating orbital elements averaged over one orbital period \textcolor{black}{as
 \eqi \left\langle\dert\Psi t\right\rangle = \left(\rp{n}{2\pi}\right)\int_0^{P_{\rm b}}\left(\dert\Psi t\right)_{\rm G}dt, \Psi=a,e,I,\Omega,\varpi,\mathcal{M},\lb{integrale}\eqf where it is intended that    \rfr{conz} is used in the integrand of \rfr{integrale} and $dt$ itself: $\left(d\Psi/dt\right)_{\rm G}$  represents the right-hand-sides of \rfr{Gauss} for the generic orbital element $\Psi$.  Note that, as usual in perturbation theory, $a,e,I,\Omega,\omega$ are kept fixed in performing the integration with respect to the fast variable $E$. Indeed, their actual variations due to several non-Keplerian effects occur over characteristic   timescales which are quite longer than $P_{\rm b}$. In the following, the brackets $\left\langle\cdots\right\rangle$ denoting the average over one orbital period will be omitted for brevity. One finally gets}
 \eqi
 \left\{
 \begin{array}{lll}
 \dert a t & = & 0, \\ \\
 \dert e t &=& -\rp{3 B c^2 \Delta Q\sqrt{1-e^2}}{2 a n}\left[
 \kz\si\co +\ci\co\left(\ky\cO-\kx\sO\right)-\right. \\ \\
 &-&\left.\so\left(\kx\cO+\ky\sO\right)\right], \\ \\
 \dert I t &=& \rp{3B c^2 \Delta Q e\co}{2a n\sqrt{1-e^2}}\left[\kz\ci+\si\left(\kx\sO-\ky\cO\right)\right], \\ \\
 \dert\Omega t & = & \rp{3B c^2 \Delta Q e\so}{2a n\sqrt{1-e^2}}\left(\kx\sO-\ky\cO+\kz\cot I\right), \\ \\
 \dert\varpi t & = & -\rp{3B c^2\Delta Q}{2e an\sqrt{1-e^2}}\left\{
 \left(e^2-1\right)\co\left(\kx\cO+\ky\sO\right)+\right.\\ \\
 &+&\left.\so\left[-\kz\si+\left(e^2-\ci\right)\left(\ky\cO-\kx\sO\right)+e^2\kz\tan\left(\rp{I}{2}\right)\right]\right\}, \\ \\
 \dert{\mathcal{M}}t &=& -\rp{3B c^2\Delta Q\left(1+e^2\right)}{2e an}\left\{
 \co\left(\kx\cO+\ky\sO\right)+\right.\\ \\
 &+&\left.\so\left[\kz\si +\ci\left(\ky\cO-\kx\sO\right)\right]
 \right\}.
 \end{array}\lb{aratan}
 \right.
 \eqf
 We remark that the expressions in \rfr{aratan} are exact in the sense that no simplifying approximations either in $e$ or in $I$ were assumed in the calculation; moreover,
 they are valid for a generic direction $\bds{\hat{k}}$ of the dipolar gradient of $\zeta$. It can be noticed that the semi-major axis remains unchanged, while the long-term variations of the inclination and the node vanish for circular orbits. The formula for $d\Omega/dt$ becomes singular for $I\rightarrow 0$; the same occurs for $d\varpi/dt$ and $d\mathcal{M}/dt$ as well for $e\rightarrow 0$. In general, the long-term changes of \rfr{aratan} are not secular trends because of the modulations introduced by the slowly time-varying orbital elements themselves occurring in real astronomical scenarios like the Earth and the Moon, and the Sun and its planets. In the calculation yielding \rfr{aratan} it was assumed that their frequencies were much smaller than the orbital one, so to keep them constant over one orbital revolution.

 The instantaneous changes of the $R-T-N$ components of the test particle's position vector $\bds r$ can be worked out from the following general expression \citep{Casotto}
 \eqi
 \left\{
\begin{array}{lll}
  \Delta R(f) &=& \left(\rp{r}{a}\right)\Delta a(f)-a \cos f\Delta e(f) + \left(\rp{a e \sin f}{\sqrt{1-e^2}}\right)\Delta \mathcal{M}(f), \\  \\
  \Delta T(f) &=& a\sin f\left[1+\rp{r}{a(1-e^2)}\right]\Delta e(f) + r[\Delta \omega(f)+\cos I\Delta\Omega(f)] +\left(\rp{a^2}{r}\right)\sqrt{1-e^2}\Delta \mathcal{M}(f), \\ \\
  \Delta N(f) &=& r\left[\sin u\Delta I(f)-\cos u\sin I\Delta\Omega(f) \right],
\end{array}\lb{RTN}
 \right.
 \eqf
 In the case of \rfr{maronna}, we have that the $R-T-N$ shifts of the position, averaged over one orbital revolution, are
  \eqi
 \left\{
 \begin{array}{lll}
 \Delta R &=& \rp{3\pi B c^2\Delta Q\sqrt{1-e^2}}{n^2}\left[\kz\co\si+\ci\co\left(\ky\cO-\kx\sO\right)-\right.\\ \\
 &-&\left.\so\left(\kx\cO+\ky\sO\right)\right], \\ \\
 \Delta T &=& -\rp{6\pi Bc^2\Delta Q}{\sqrt{1-e^2}n^2}\left\{
 \co\left(\kx\cO+\ky\sO\right)+\right.\\ \\
 &+&\left.\so\left[\kz\si + \ci\left(\ky\cO-\kx\sO\right)\right]\right\}, \\ \\
 \Delta N & = & 0.
 \end{array}\lb{drdtdn}
 \right.
 \eqf
 Also the expressions of \rfr{drdtdn} are exact in both $e$ and $I$; notice also that they   present no singularities for both $e\rightarrow 0$ and $I\rightarrow 0$. Moreover, they, in general,  vanish neither for circular orbits nor for $I=0$.

For the instantaneous $R-T-N$ perturbations of the velocity vector $\bds v$ we have, in general, \citep{Casotto}
 \eqi
 \left\{
 \begin{array}{lll}
  \Delta v_{R}(f) &=& -\rp{n\sin f}{\sqrt{1-e^2}}\left[\rp{e}{2}\Delta a(f)+\rp{a^2}{r}\Delta e(f)\right]-\rp{n a^2\sqrt{1-e^2}}{r}\left[\Delta\omega(f)+\cos I\Delta\Omega(f)\right]-\left(\rp{n a^3}{r^2}\right)\Delta \mathcal{M}(f), \\  \\
  \Delta v_{T}(f) &=& -\left(\rp{na\sqrt{1-e^2}}{2r}\right)\Delta a(f)+\rp{na(e+\cos f)}{(1-e^2)^{3/2}}\Delta e(f)+\rp{nae\sin f}{\sqrt{1-e^2}}\left[\Delta\omega(f)+\cos I\Delta\Omega(f)\right], \\ \\
  \Delta v_{N}(f) &=& \rp{na}{\sqrt{1-e^2}}\left[\left(\cos u+e\cos\omega\right)\Delta I(f) + \left(\sin u + e\sin\omega\right)\sin I\Delta\Omega(f)\right].
\end{array}
\lb{vRTN}
 \right.
 \eqf
In the case of \rfr{maronna}, \rfr{vRTN} yields \textcolor{black}{the following expressions for the velocity changes over one orbital revolution}
\eqi
 \left\{
 \begin{array}{lll}
 \Delta v_R & = & \rp{3\pi Bc^2 \Delta Q\left[1+e\left(2-e\right)\right]}{\left(1-e^2\right)n}\left\{
 \co\left(\kx\cO+\ky\sO\right) +\right. \\ \\
 &+&\left.\so\left[\kz\si + \ci\left(\ky\cO-\kx\sO\right)\right]\right\}, \\ \\
  \Delta v_T & = & -\rp{3\pi Bc^2 \Delta Q}{\left(1-e\right)n}\left[\kz\co\si +\ci\co\left(\ky\cO-\kx\sO\right)-\right.\\ \\
  &-&\left.\so\left(\kx\cO+\ky\sO\right)\right], \\ \\
  \Delta v_N & = & \rp{3\pi Bc^2 \Delta Q e}{\left(1-e\right)n}\left[\kz\ci +\si\left(\kx\sO-\ky\cO\right)\right].
 \end{array}\lb{strunz}
 \right.
 \eqf
 Also the expressions of \rfr{strunz} are exact in both $e$ and $I$, and are not singular for any particular value of them. Notice that $\Delta v_N$ vanishes for circular orbits.
\section{Confrontation with the observations}\lb{confronto}
 For the sake of definiteness, in the following we will focus on the fine structure constant whose spatial variation has the most stringent empirical support  so far.

 \textcolor{black}{The following general considerations are in order to properly contextualize and understand the content of the next sections. Our aim is to evaluate as realistically as possible the magnitude of the predicted anomalous effects  on quantities which are empirically determined from observations by astronomers. It is important to note that they were produced by adopting  force models including all the standard known Newtonian/Einsteinian dynamics which, thus, did not include the dipolar effects considered here. In principle, such observationally determined quantities  should account for them, but it is quite possible that in the estimation procedure of, say, the planetary state vectors the exotic signatures were partially or totally absorbed, especially if their magnitude is small with respect to the accuracy of the orbit determination process. Thus, the subsequent comparison of the theoretically computed Stark-like effects  with the present-day empirical accuracies  is just a preliminary, although quite reasonable and necessary, step to investigate if the putative orbital anomalies considered here may be considered as potentially measurable in further, dedicated analyses. We do not intent to put on the test their actual measurability: indeed, to this aim it would be necessary to reprocess the entire lunar/planetary data sets with ad-hoc modified dynamical models explicitly including the Stark-type effect of interest, and dedicated solve-for parameter(s) should be estimated in the new global fit. Of course, such a non-trivial task would be worthwhile only if there are some hopes that the effects one is interested in are, at least indicatively, detectable.    }
 \subsection{The planets}
 In order to exploit the latest results in the field of  planetary orbit determination, in Table \ref{tavola} we quote the values of $\Delta Q$ for several Sun-planets pairs. We computed them by means of eq. (42) in \citet{DamDon}, based on earlier results \citep{DamDon010a,DamDon010b}. As it can be expected, $\Delta Q$ is larger for the rocky planets, mostly made of iron cores and silicate mantels, with respect to the gaseous giants like Saturn, for which accurate orbital data are now available from the Cassini mission \citep{Fie010,FiengaJournees010,pitjeva}, whose composition \citep{Sat} is more similar to that of the Sun. For our parent star we assumed that hydrogen and helium constitute $74\%$ and $25\%$ of its total mass, respectively \citep{Sun}.
 Concerning Mercury, it is known since a long time \citep{Ure1,Ure2} that its iron-to-silicate ratio $\lambda$ must be much larger than that of any other terrestrial planet and the Moon; for it we assumed $\lambda=70:30$ \citep{Mercu}. Venus has a rather similar composition with respect to the Earth \citep{plane}, for which we assume $\lambda=32:68$ \citep{plane}.
 According to a spacecraft data-based analysis by \citet{Yoder}, the venusian iron/silicate ratio is\footnote{Indeed, according to Figure 2 by \citet{Yoder}, the mean core density of Venus is $\overline{\varrho}_c=10.16$ g cm$^{-3}$ for a value of the core radius of $R_c=3100$ km.} $\lambda=26:74$. Mars is an intermediate case between the Earth and the Moon \citep{plane}, which is almost entirely made of silicates \citep{WilBog}; latest determinations from orbiting probes \citep{Folk} allow to infer\footnote{Cfr. with Figure 26 of \citep{Folk}. We used  $\varrho_c=6.7$ g cm$^{-3}$ for the core density  and  $R_c=1680$ km for the core radius \citep{Khan}.}  $\lambda=21:79$ for the red planet. For Saturn we take a mass fraction for helium of the order of $0.13$ from the latest Cassini-based measurements \citep{Sat}, while the rest is assumed to be made mostly of hydrogen.
 \begin{table*}[ht!]
\caption{Values  $\Delta Q\doteq Q_{\rm p}-Q_{\odot}$ for various planet-Sun pairs inferred from eq. (36) of \citet{DamDon} for the case of the fine structure constant. See the discussion in the text for the details concerning the composition of the various bodies.
}\label{tavola}
\centering
\bigskip
\begin{tabular}{lllll}
\hline\noalign{\smallskip}
Mercury & Venus & Earth & Mars & Saturn  \\
\noalign{\smallskip}\hline\noalign{\smallskip}
$1.979\times 10^{-3}$ & $1.539\times 10^{-3}$ & $1.599\times 10^{-3}$ & $1.489\times 10^{-3}$ & $-2.8\times 10^{-5}$ \\
\noalign{\smallskip}\hline\noalign{\smallskip}
\end{tabular}
\end{table*}
\subsubsection{The perihelion precessions}
By using the figures of Table \ref{tavola} and the result of \rfr{aratan}, assumed to be valid in a heliocentric reference frame with the mean ecliptic and equinox at the epoch J$2000.0$,  we compute the long-term perihelion precessions induced by \rfr{maronna}. We adopt \rfr{kappa} for $\bds{\hat{k}}$ and \rfr{fina} for $B$.  Table \ref{peritavola} displays our results in milliarcseconds per century (mas cty$^{-1}$).
\begin{table*}[ht!]
\caption{Perihelion precessions $\dot\varpi$, in milliarcseconds per century (mas cty$^{-1}$), for the four inner planets and Saturn induced by a dipole-type gradient of the fine structure constant according to \rfr{aratan}. We used \rfr{kappa} for $\bds{\hat{k}}$, \rfr{fina} for $B$, and the figures of Table \ref{tavola} for $\Delta Q$. The  Keplerian orbital elements of all the planets were kept fixed to their values referred to the mean ecliptic and equinox at the epoch J2000.0: they were retrieved from the NASA WEB interface HORIZONS (Author: J. Giorgini. Site Manager: D. K. Yeomans.
Webmaster: A. B. Chamberlin).
}\label{peritavola}
\centering
\bigskip
\begin{tabular}{lllll}
\hline\noalign{\smallskip}
 Mercury  & Venus & Earth & Mars & Saturn  \\
\noalign{\smallskip}\hline\noalign{\smallskip}
 $-0.0016$ & $-0.0369$ & $-0.0244$ & $0.0017$ & $0.0004$ \\
\noalign{\smallskip}\hline\noalign{\smallskip}
\end{tabular}
\end{table*}

They should be compared with the latest empirical determinations of the corrections $\Delta\dot\varpi$ to the standard planetary perihelion precessions obtained by fitting  accurate dynamical force models, which include most of the standard Newtonian and Einsteinian dynamical effects, to observational data records spanning about one century. They are resumed in Table \ref{perirate}.
\begin{table*}[ht!]
\caption{Estimated corrections $\Delta\dot\varpi$, in milliarcseconds per century (mas cty$^{-1}$), to the standard Newtonian-Einsteinian secular precessions of the longitudes of the perihelia $\varpi$ of the eight planets plus Pluto determined with the EPM2008 \protect{\citep{pitjeva}}, the INPOP08 \protect{\citep{Fie010}},  and the INPOP10a \protect{\citep{FiengaJournees010}} ephemerides. Only the usual Newtonian-Einsteinian dynamics was modelled, so that, in principle, the corrections $\Delta\dot\varpi$ \textcolor{black}{may} account for any other unmodelled/mismodelled dynamical effect. Concerning the values quoted in the third column from the left, they correspond to the smallest uncertainties reported by \protect{\citet{Fie010}}. Note the small uncertainty in the correction to the precession of the terrestrial perihelion, obtained by processing Jupiter VLBI data \protect{\citep{Fie010}}.
}\label{perirate}
\centering
\bigskip
\begin{tabular}{llll}
\hline\noalign{\smallskip}
Planet & $\Delta\dot\varpi$  \protect{\citep{pitjeva}}  & $\Delta\dot\varpi$  \protect{\citep{Fie010}} &  $\Delta\dot\varpi$  \protect{\citep{FiengaJournees010}} \\
\noalign{\smallskip}\hline\noalign{\smallskip}
Mercury & $ -4 \pm 5 $  & $ -10\pm 30$ & $ 0.2\pm 3$ \\
Venus & $ 24\pm 33$  & $-4\pm 6 $ & $ - $ \\
Earth & $ 6\pm 7$  & $ 0 \pm 0.016 $ & $ - $\\
Mars & $ -7\pm 7$  & $0\pm 0.2 $ & $ - $\\
Jupiter & $ 67\pm 93$  & $142\pm 156$ & $ - $\\
Saturn & $ -10\pm 15$ & $-10\pm 8$ & $ 0\pm 2$ \\
Uranus & $ -3890\pm 3900$  & $0\pm 20000$ & $ - $\\
Neptune & $ -4440\pm 5400 $  & $0\pm 20000$ & $ - $\\
Pluto & $ 2840 \pm 4510 $  & $-$ & $ - $\\
\noalign{\smallskip}\hline\noalign{\smallskip}
\end{tabular}
\end{table*}

It can be noticed that the predicted precessions of Table \ref{peritavola} are, in general, smaller than the present-day uncertainties in estimating $\Delta\dot\varpi$ by $2-3$ orders of magnitude, apart from the value obtained by \citet{Fie010} for the Earth's perihelion rate by including some VLBI points for Jupiter.
\subsubsection{The Earth-planet ranges}
Since a direct, unambiguous observable is the range $\rho$ between the Earth and a  planet\footnote{\textcolor{black}{It is particularly true when a spacecraft specially outfitted for ranging measurements orbits the planetary target of interest.}}, and in view of the possible future implementation of  interplanetary laser ranging facilities accurate to  cm level \citep{plr1,plr2,plr3}, we investigate the impact that a violation of the equivalence principle due to a Stark-type acceleration like  \rfr{maronna} may have on such a quantity. We do not consider its consequences on the propagation of the electromagnetic waves involved in ranging.

Figure \ref{figura} depicts the numerically produced range perturbations $\Delta\rho$ induced by \rfr{maronna} for Mercury, Venus, Mars and Saturn over different time spans.
\begin{figure*}[ht!]
\centering
\begin{tabular}{cc}
\epsfig{file=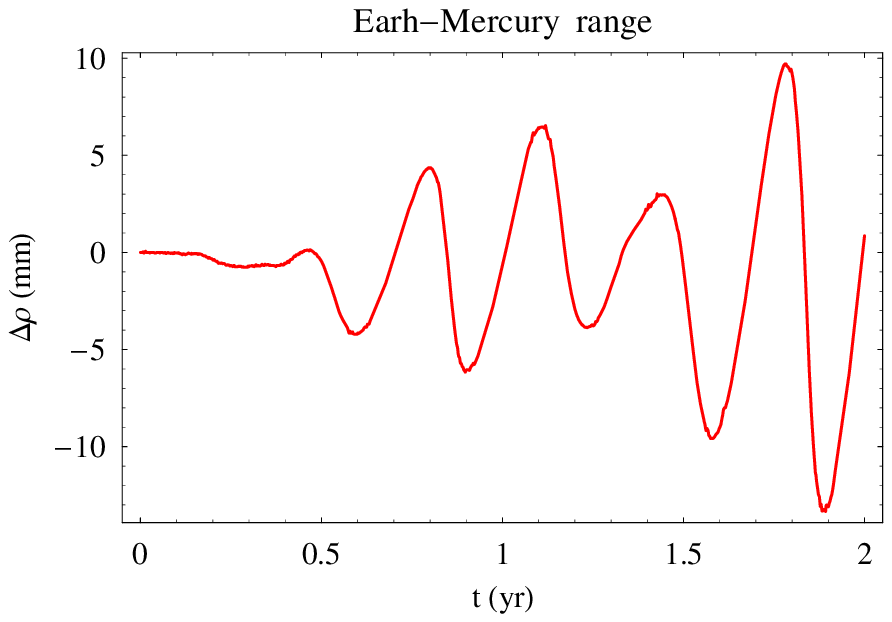,width=0.30\linewidth,clip=} & \epsfig{file=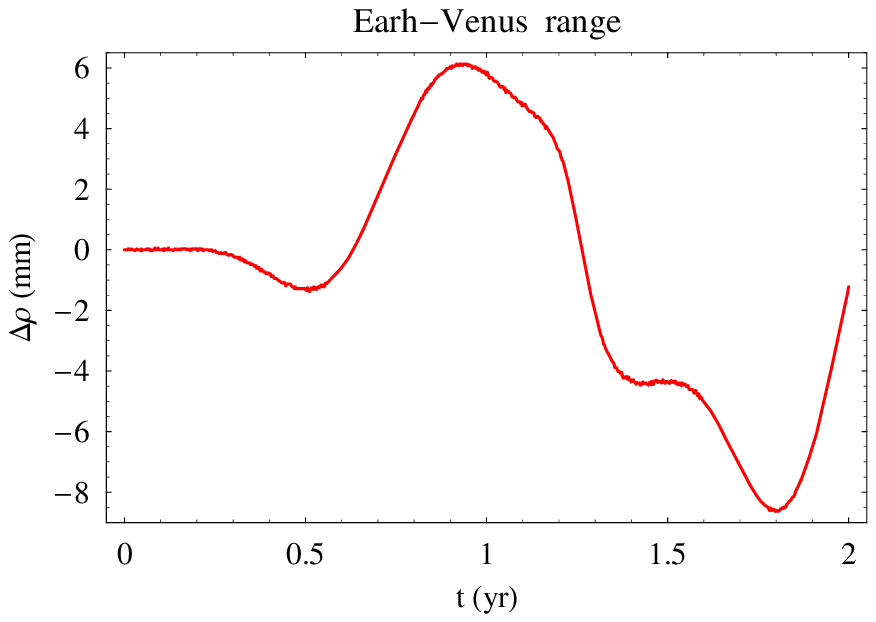,width=0.30\linewidth,clip=}  \\
\epsfig{file=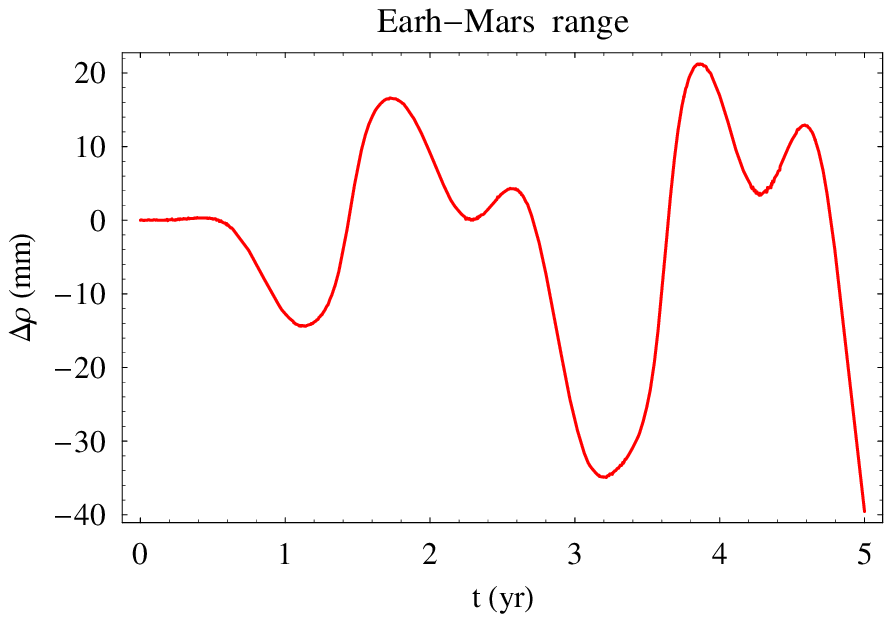,width=0.30\linewidth,clip=} & \epsfig{file=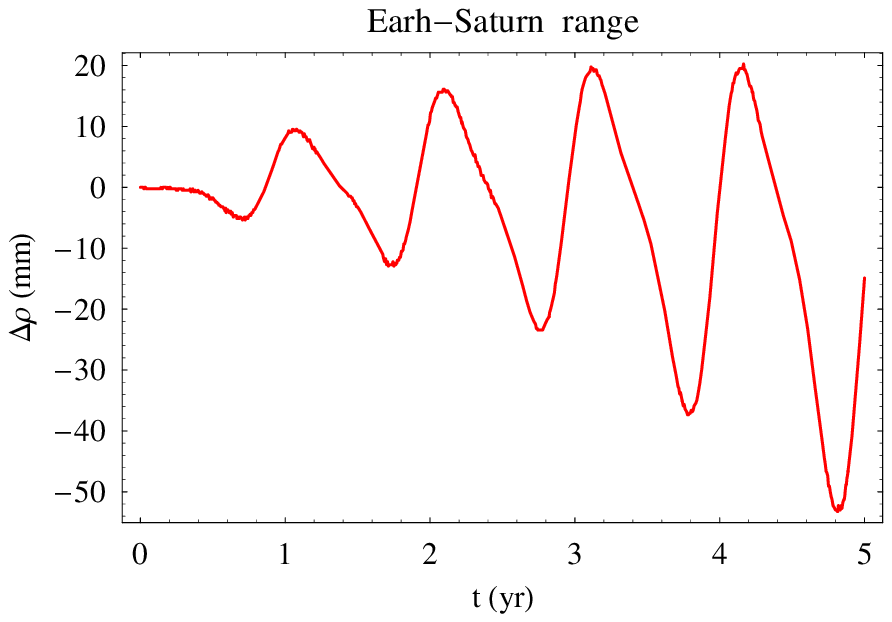,width=0.30\linewidth,clip=}
\end{tabular}
\caption{Numerically produced Earth-planet range perturbations $\Delta\rho$, in mm, induced by the Stark-like extra-acceleration of \rfr{maronna} for Mercury, Venus, Mars and Saturn. We adopted the figures of Table \ref{tavola} for $\Delta Q$, and \rfr{fina} and \rfr{kappa} for $B$ and $\bds{\hat{k}}$, respectively. Each curve is the difference between the  time series for a given Earth-planet range $\rho$ computed by numerically integrating the equations of motion in cartesian coordinates with and without \rfr{maronna}. Both the integrations share the same initial conditions retrieved from the WEB interface HORIZONS (Author: J. Giorgini. Site Manager: D. K. Yeomans. Webmaster: A. B. Chamberlin) by JPL, NASA. They refer to a heliocentric frame with the mean equinox and ecliptic  at  the epoch J$2000.0$. The \textcolor{black}{total ranges of} integration are $\Delta t=2$ yr for Mercury and Venus, and $\Delta t=5$ yr for Mars and Saturn.
}\lb{figura}
\end{figure*}
The quantitative features of the signatures of Figure \ref{figura} are resumed in Table \ref{resume}.
\begin{table*}[ht!]
\caption{Peak-to-peak maximum amplitude $|\Delta\rho|^{\rm max}$, mean $\left\langle\Delta\rho\right\rangle$ and variance $\sigma_{\Delta\rho}$, in mm, of the range signals of Figure \ref{figura} caused by the  Stark-like acceleration of \rfr{maronna} for the spatial gradient fine structure constant. The \textcolor{black}{total ranges of} integration are $\Delta t=2$ yr for Mercury and Venus, and $\Delta t=5$ yr for Mars and Saturn. }
\label{resume}
\centering
\bigskip
\begin{tabular}{llll}
\hline\noalign{\smallskip}
Planet & $|\Delta\rho|^{\rm max}$ (mm) & $\left\langle\Delta\rho\right\rangle$ (mm)&  $\sigma_{\Delta\rho}$ (mm)  \\
\noalign{\smallskip}\hline\noalign{\smallskip}
Mercury & $23.1$ & $-0.9$ & $4.4$ \\
Venus & $14.8$ & $-0.8$ & $4.2$\\
Mars & $60.8$ & $-2.4$ & $14.7$\\
Saturn & $73.7$ & $-5.1$ & $16.2$ \\
\noalign{\smallskip}\hline\noalign{\smallskip}
\end{tabular}
\end{table*}
It can be noticed that, although their temporal signatures are rather distinct,  their magnitudes are quite small, amounting to about $0.1-10$ mm.
\subsection{The Moon}
In Table \ref{tavolaluna} we quote the relevant orbital and physical parameters of the Earth-Moon system.
\begin{table*}[ht!]
\caption{Relevant physical and {osculating} orbital parameters of the Earth-Moon system. $a$ is the semi-major axis. $e$ is the eccentricity. The inclination $I$ {is referred} to the mean ecliptic at J2000.0. $\Omega$ is the longitude of the ascending node, it circulates with a period of $6798.38$ d, and is referred to the mean equinox and ecliptic at J2000.0. $\omega$ is the argument of pericenter: its period is $2191.50$ d. $G$ is the Newtonian gravitational constant.  The masses of the Earth and the Moon are $M$ and $m$, respectively. $\Delta Q\doteq Q_{\leftmoon}-Q_{\oplus}$ refers to the fine structure constant: its value comes from \protect{\citet{DamDon}}. The orbital parameters of the Moon were retrieved from the WEB interface HORIZONS (Author: J. Giorgini. Site Manager: D. K. Yeomans.
Webmaster: A. B. Chamberlin) by JPL, NASA, at  the epoch J2000.0.
}\label{tavolaluna}
\centering
\bigskip
\begin{tabular}{llllllll}
\hline\noalign{\smallskip}
$a$ (m) & $e$ & $I$ (deg) & $\Omega$ (deg) & $\omega$ (deg) & $GM$ (m$^3$ s$^{-2}$) & $m/M$  & $\Delta Q$\\
\noalign{\smallskip}\hline\noalign{\smallskip}
$3.81219\times 10^8$ & $0.0647$ & $5.24$ & $123.98$ & $-51.86$ & $3.98600\times 10^{14}$ &  $0.012$ & $-3.2\times 10^{-4}$\\
\noalign{\smallskip}\hline\noalign{\smallskip}
\end{tabular}
\end{table*}
Table \ref{tavolalunaeffetti} displays some computed orbital effects of a dipole-type spatial gradient of the fine structure constant for the Moon: a geocentric frame with the mean equinox and ecliptic at the epoch J$2000.0$ is assumed in applying \rfr{aratan}, \rfr{drdtdn} and \rfr{strunz}.
\begin{table*}[ht!]
\caption{Orbital effects caused by \rfr{maronna} on the Moon in the case of a dipolar-like gradient of the fine structure constant. For the Earth-Moon system the figures of Table \ref{tavolaluna} were used. The variation of the eccentricity and the perigee precession were worked out from \rfr{aratan}. The position and velocity perturbations were computed according to \rfr{drdtdn} and \rfr{strunz}, respectively. All the effects are averaged over one Moon's orbital revolution. The present-day accuracies in determining the lunar eccentricity and perigee secular variations
 are of the order of $3\times 10^{-12}$ yr$^{-1}$ \protect{\citep{WilBog}} and $10^{-1}$ mas yr$^{-1}$ \protect{\citep{Wi,Mull1,Mull2}}, respectively, while the Earth-Moon post-fit range residuals are consistently approaching the mm level \protect{\citep{Will,Mur,Batt}}.
}\label{tavolalunaeffetti}
\centering
\bigskip
\begin{tabular}{lllllll}
\hline\noalign{\smallskip}
$\dot e$ $\left(\rp{1}{\rm yr}\right)$ & $\dot\varpi$ $\left(\rp{\rm mas}{\rm yr}\right)$ & $\Delta R$ (mm)  & $\Delta T$ (mm) & $\Delta v_R$ $\left(\rp{\rm mm}{\rm yr}\right)$ & $\Delta v_T$ $\left(\rp{\rm mm}{\rm yr}\right)$ & $\Delta v_N$ $\left(\rp{\rm mm}{\rm yr}\right)$\\
\noalign{\smallskip}\hline\noalign{\smallskip}
$-3\times 10^{-14}$ & $3.5\times 10^{-4}$ & $8\times 10^{-4}$ & $-6.3\times 10^{-3}$ & $0.32$ & $-0.07$ & $0.01$ \\
\noalign{\smallskip}\hline\noalign{\smallskip}
\end{tabular}
\end{table*}

It may be interesting to notice that the magnitude of the long-term variation of the eccentricity is $10^{-14}$ yr$^{-1}$: this implies that the gradient of the fine structure constant cannot be the cause of the anomalous secular increase of the lunar eccentricity \citep{JGR,hjk,WilBog} $\dot e=(9\pm 3)\times 10^{-12}$ yr$^{-1}$ recently discussed in literature \citep{And010,Iorio}. The predicted perigee precession is as large as  $3.3\times 10^{-4}$ mas yr$^{-1}$; the present-day accuracy in determining the lunar orbital precessions is at  $0.1$ mas yr$^{-1}$ level \citep{Wi,Mull1,Mull2}. According to Table \ref{tavolalunaeffetti}, the radial and transverse position shifts per orbit are of the order of $\mu$m. The radial variation per orbit of the velocity amounts to about $0.1$ mm yr$^{-1}$, while the magnitudes of the other two components are one order of magnitude smaller. Improvements in both  measurement techniques and in dynamical modeling are pointing towards post-fit Earth-Moon range residuals at mm level \citep{Will,Mur,Batt}. The peculiar temporal patterns of the signals investigated may help in separating them from other dynamical standard effects if and when the level of accuracy required to determine them will eventually be reached.

\textcolor{black}{We  numerically integrated the Earth-Moon range over $\Delta t=30$ yr;}
%
%
%
%
%
%
%
%
%
\textcolor{black}{its principal quantitative features are reported in Table \ref{resumeluna}.}
\begin{table*}[ht!]
\caption{Peak-to-peak maximum amplitude $|\Delta\rho|^{\rm max}$, mean $\left\langle\Delta\rho\right\rangle$ and variance $\sigma_{\Delta\rho}$, in mm, of the \textcolor{black}{numerically integrated} geocentric lunar range signal
caused by the  Stark-like acceleration of \rfr{maronna} for the spatial gradient fine structure constant. \textcolor{black}{It was produced as for the planets: the}  \textcolor{black}{total ranges of the} integrations are$\Delta t=30$ yr. }
\label{resumeluna}
\centering
\bigskip
\begin{tabular}{lll}
\hline\noalign{\smallskip}
 $|\Delta\rho|^{\rm max}$ (mm) & $\left\langle\Delta\rho\right\rangle$ (mm)&  $\sigma_{\Delta\rho}$ (mm)  \\
\noalign{\smallskip}\hline\noalign{\smallskip}
$4.5$ & $3\times 10^{-4}$ & $0.8$  \\
\noalign{\smallskip}\hline\noalign{\smallskip}
\end{tabular}
\end{table*}
While the maximum peak-to-peak amplitude is as large as a few mm, its mean amounts to $10^{-4}$ mm, with a variance of $0.8$ mm. They are certainly quite small figures, even for a so large time span as the one adopted.
\section{Summary and conclusions}\lb{conclusioni}
We considered a dipolar spatial variation of the fine structure constant, for which  empirical evidence at about $4\sigma$ level currently exists, and looked at the impact that the resulting Stark-like anomalous acceleration may have on the orbital motion of a test particle around a central body. Since the relative two-body Stark-type acceleration depends in a certain way on the different composition of the bodies involved, it violates the equivalence principle.

We, first, analytically worked out the  long-term, i.e. averaged over one orbital revolution, variations of all the six osculating Keplerian orbital elements of the test particle. We did not restrict ourselves to any specific orbital configuration; moreover, we made no \textit{a-priori} assumptions on the fixed direction of the gradient of the fine structure constant. Thus, our results are quite general. It turned out that, apart from the semi-major axis, all the other osculating Keplerian orbital elements experience non-vanishing long-term variations which, in real astronomical scenarios, would be modulated primarily by the low frequencies of the inclination, the node and the pericenter due to the standard mutual N-body interactions. Then, we worked out the changes per orbit of the position and velocity vectors of the test particle. Also in this case, the expressions obtained are valid for any orbital geometry and gradient direction. Only  the long-term shift of the out-of-plane component of the position vector vanishes, while the other non-vanishing ones exhibit characteristic slow time-varying modulations because of the presence of the inclination, the node and the pericenter in their expressions.

Subsequently, we computed our predicted effects for some Sun-planet pairs of the solar system and for the Earth-Moon system. In particular, the perihelion precessions of the inner planets, which exhibit the largest Stark-like accelerations because of their markedly different composition with respect to the Sun, are of the order of $10^{-2}-10^{-3}$ mas cty$^{-1}$. We numerically computed times series some years long of the Stark-induced  perturbation of the interplanetary range for some  Earth-planet pairs as well; their magnitude is approximately $0.1-10$ mm. Concerning the Moon, the long-term precession of its perigee is of the order of $10^{-4}$ mas yr$^{-1}$, while the long-term variation of its eccentricity is a decrease at the  $10^{-14}$ yr$^{-1}$ level. The radial and transverse shifts per orbit of the Moon's position amount to some $\mu$m, while the radial, transverse and normal variations per orbit of its velocity are of the order of $0.1-0.01$ mm yr$^{-1}$. In addition to that, we also numerically produced an Earth-Moon perturbed range time series spanning 30 yr: its main quantitative features are at a sub-mm level, with a nominal peak-to-peak amplitude of a few mm and a mean of $0.3$ $\mu$m.

The level of accuracy in empirically determining the corrections to the standard Newtonian-Einsteinian perihelion precessions from planetary observations is currently $2-3$ orders of magnitude larger than the predicted Stark-like effects, apart from, perhaps, the Earth provided that VLBI data are included in the data processing. Recent years have seen increasing efforts toward the implementation of the planetary
laser ranging  technique, which should be accurate at cm level. Concerning the Moon, its orbital precessions can be presently determined at a $10^{-1}$ mas yr$^{-1}$ level of accuracy, while the LLR post-fit range residuals are approaching the mm level.

In conclusion, detecting the putative spatial variations of the fine structure constant through its orbital effects on the major bodies of the solar system  seems unlikely or very difficult in the near future.

\end{document}